\documentclass[%
 reprint,
 amsmath,amssymb,
 aps,
]{revtex4-2}

\usepackage{graphicx}% Include figure files
\usepackage{dcolumn}% Align table columns on decimal point
\usepackage{bm}% bold math
\begin{document}

\preprint{APS/123-QED}

\title{High-frequency, resonant acousto-optic modulators fabricated in a MEMS foundry platform}% Force line breaks with

\author{Stefano Valle}
 
\author{Krishna C. Balram}%
 \email{krishna.coimbatorebalram@bristol.ac.uk}
\affiliation{%
 Quantum Engineering Technology Labs, H. H. Wills Physics Laboratory and Department of Electrical and Electronic Engineering, University of Bristol, Bristol BS8 1UB, United Kingdom
}%

\date{\today}% It is always \today, today,

\begin{abstract}
We report the design and characterization of high frequency, resonant acousto-optic modulators (AOM) in a MEMS foundry process. The doubly-resonant cavity design, with short ($L{\sim}10.5\, {\mu}m$) acoustic and optical cavity lengths, allows us to measure acousto-optic modulation at GHz frequencies with high modulation efficiency. In contrast to traditional AOMs, these devices rely on the perturbation induced by the displacement of cavity boundaries, which can be significantly enhanced in a suspended geometry. This platform can serve as the building block for fast 2D spatial light modulators (SLM), low-cost integrated free space optical links and optically enhanced low-noise RF receivers.
\end{abstract}

\maketitle

%\tableofcontents

\section{Introduction}
Acousto-optic (AO) devices \cite{Korpel1981Acousto-OpticsAFundamentals} exploit the interaction between light and sound waves to enable a wide variety of signal processing functions, in particular optical modulation, spectral filtering and frequency shifting. While acousto-optic modulators (AOMs) cannot match the electrical bandwidths of state-of-the-art travelling wave electro-optic modulators (EOMs)  \cite{Wang2018IntegratedVoltages}, they offer a significant efficiency advantage. AOMs are more efficient than EOMs at a given modulation frequency  because the acoustic wave amplitude in an AOM can be resonantly enhanced to far greater levels than the corresponding radio frequency (RF) electric field in an EOM. For example, an acoustic cavity at high frequencies (>1 GHz) can easily have a mechanical quality factor ($Q_{mech}$) greater than 1000 whereas it is very hard to engineer compact RF cavities with a quality factor greater than 100. In addition, the acoustic and optical fields have comparable wavelengths ($\lambda_{Si}^{acoustic}(5.6\,GHz)\approx1.55 {\mu}m$) which allows us to confine them in nanoscale cavities and increase the interaction strength by engineering larger spatial mode overlap \cite{safavi2019controlling}. This becomes critical when long interaction lengths are not available, e.g. in surface-normal operation. To exploit this efficiency advantage of AOMs, two key challenges must be overcome: low operation frequencies and foundry (CMOS) incompatibility. AOMs have traditionally been built with bulk crystals like quartz and tellurium dioxide \cite{Valle2015ImagingMid-IR}, which are not available in foundries. Moreover, traditional AOMs use relatively thick (~ 10 ${\mu}$m) piezoelectric crystals as transducers which limit the operating frequency to $\approx500$ MHz.  By switching to a piezoelectric thin film on silicon platform, one can trade-off a lower acousto-optic figure of merit for the benefits of higher frequency operation and large-scale integration.

An efficient high-frequency AOM can serve as the building block for a wide variety of applications. For example: a 2D array of these devices can be used as a high-speed spatial light modulator with tens of MHz bandwidth, far exceeding those of liquid crystal on silicon devices \cite{Johnson1993SmartSilicon}. By increasing the modulation bandwidth to $\approx$ 100 MHz, we can envision a viable free space optical interconnect link between processors and memory that meets the stringent energy and bandwidth constraints and removes the need for an on-chip serializer/deserializer (SERDES) \cite{Siddiqui:16,miller2017attojoule}. In addition, an efficient modulator is a key component for optical approaches to RF detection \cite{Sorensen2014OpticalTransducer} and transduction, which has applications ranging from radio astronomy and MRI to quantum information processing \cite{Takeda:18,Andrews2014BidirectionalLight}. In this work, we demonstrate the operation of high frequency AOMs using a commercial MEMS foundry platform (Piezo-MUMPS).
\begin{figure}[thbp]
    \centering
    \includegraphics[width = \linewidth]{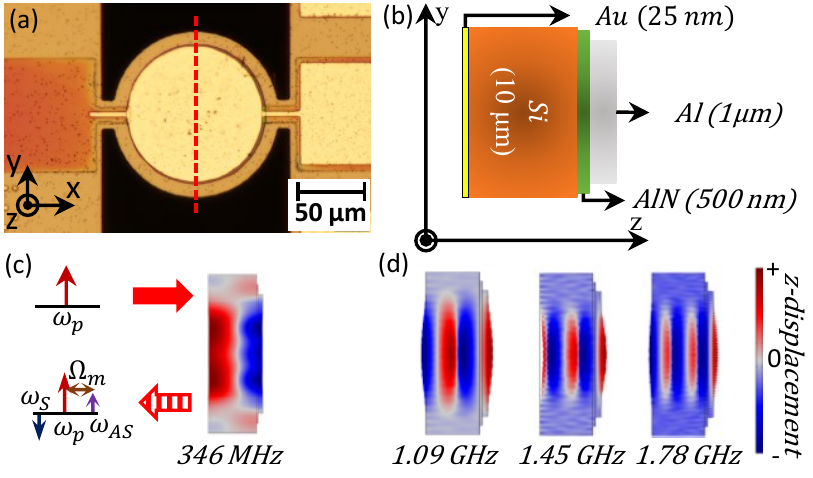}
    \caption{(a) Microscope image of a fabricated device (b) Cross-section showing layer thickness (c)   Illustration of device operation. The excitation of the resonance (at $f_{res} \approx$ 346 MHz) modifies the cavity length, producing a modulated reflected signal.(d) $z$-displacement profiles for three successive resonant modes of the cavity, calculated using COMSOL.}
    \label{fig:1}
\end{figure}

\section{Device Description and Operation} \label{sec:Device description}

The device geometry (Fig.\ref{fig:1}(a)) consists of a 500 nm piezoelectric aluminum nitride (AlN) film, sandwiched between a top electrode made of aluminum ($t_{Al} =1\,{\mu}m$) and a bottom electrode made of p-doped Si ($t_{Si} = 10\,{\mu}m$). A schematic of the device cross-section is shown in  (Fig.\ref{fig:1}(b)). The device is operated as a collinear doubly-resonant (acoustic and optical) cavity with the light incident from the silicon side (Fig.\ref{fig:1}(c)). Since the AO interaction efficiency is proportional to the intra-cavity acoustic power and the sideband amplitude scales with the slope of the optical cavity transfer function ($\sim Q_{opt}$), the doubly-resonant geometry helps to significantly enhance the interaction strength, resulting in higher modulation efficiency 
 \cite{psarobas2010enhanced,xiong2013cavity,bochmann2013nanomechanical,Balram2016CoherentCircuits,ghosh2016laterally}. In contrast to traditional AOMs that rely on the AO effect, utilizing the refractive index change induced by an acoustic wave, these devices primarily exploit the change in the optical cavity length caused by the excitation of the acoustic mode, henceforth referred to as the moving boundary effect \cite{balram2014moving}. The benefits of working with a suspended MEMS platform are immediately apparent. As both top and bottom surfaces are free to move, we can achieve larger cavity displacements for the same circulating acoustic power. In addition, the small cavity size ($L_{opt}$) allows us to work with higher frequency acoustic modes and achieve larger AO coupling strengths as the optical frequency shift scales inversely with cavity length ($d\omega_{opt}/dL_{opt}\sim\omega_{opt}/L_{opt}$). This is in contrast to traditional thin-film AOMs, which replace the piezoelectric transducer crystal with a thin film to achieve higher operation frequencies, but still rely on the elasto-optic effect in a bulk crystal or fiber to achieve modulation \cite{hickernell1988characteristics}. 

\begin{figure}[htbp]
    \centering
    \includegraphics[width = \linewidth]{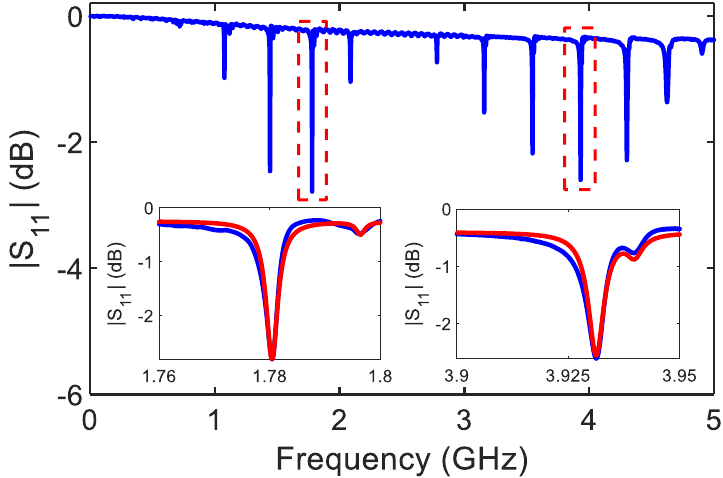}
    \caption{RF (mag.) reflection spectrum ($|S_{11}|$) of a representative device. The dips correspond to successive longitudinal acoustic cavity modes. The insets show narrowband $|S_{11}|$ spectra of two resonant modes, fit with a Lorentzian lineshape (red).}
    \label{fig:2_S11}
\end{figure}

\begin{figure}[htbp]
    \centering
    \includegraphics[width = \linewidth]{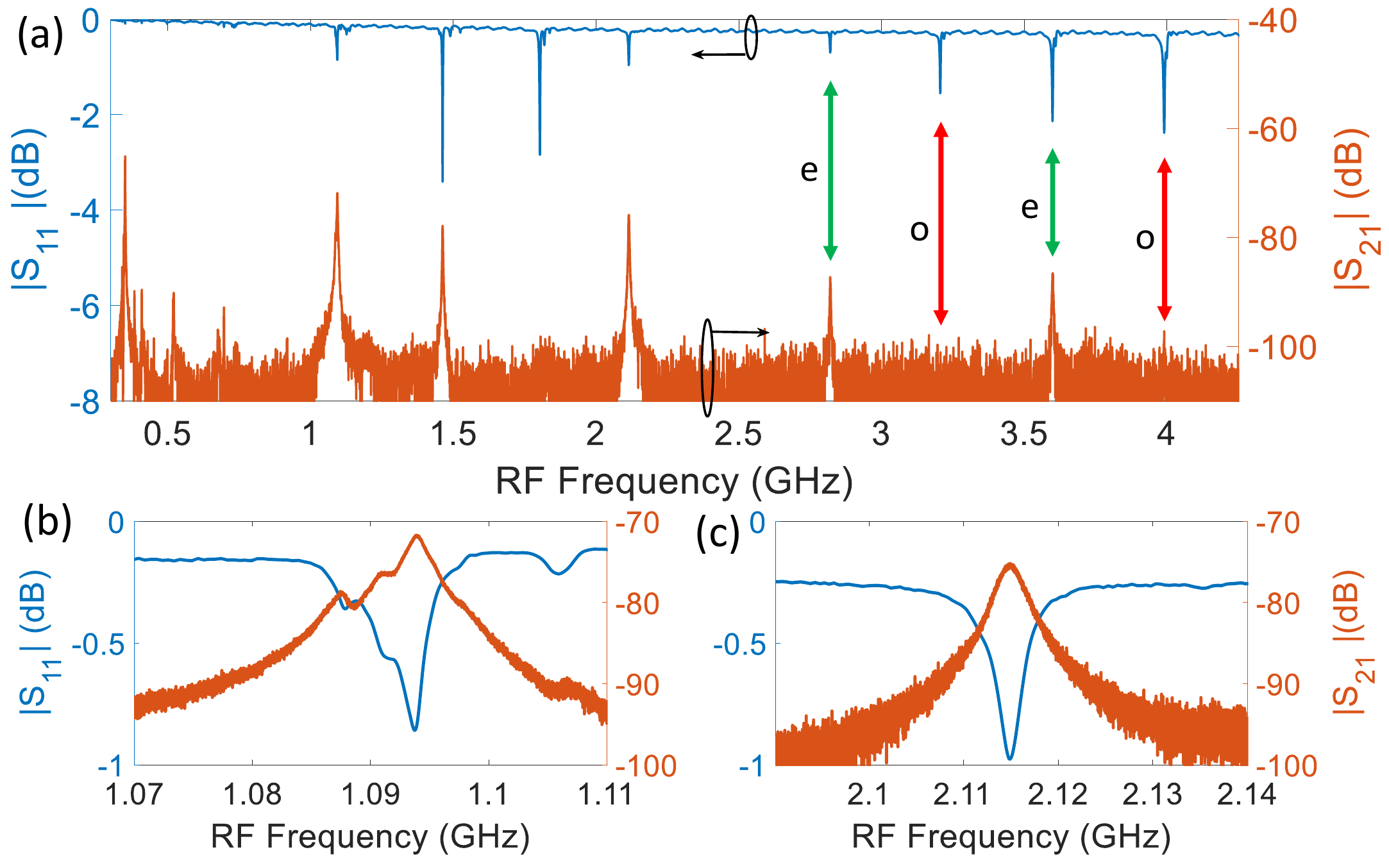}
    \caption{(a) RF reflection ($|S_{11}|$) and transmission  ($|S_{21}|$) spectra of a device showing correspondence between reflection minima and transmission maxima. The even (green) and odd (red) modes are indicated.Narrow band RF spectra of the (b) 1.096 GHz and (c) 2.115 GHz modes.}
    \label{fig:4_S11_S21}
\end{figure}

The mechanical modes of the structure are high-overtone bulk acoustic wave resonances (HBAR) \cite{lakin1993high} that are excited when an RF signal is applied across the piezoelectric AlN layer. This allows us to exploit the large $d_{33}$ coefficient of AlN for efficient RF to mechanical transduction, a process at the heart of RF filtering in modern smartphones \cite{Ruby2015APhones}.  In contrast to surface acoustic wave based acousto-optic devices \cite{li2015nanophotonic}, HBAR provides a better route for achieving high mechanical frequencies (> 10 GHz) as the transducer performance does not degrade significantly \cite{Siddiqui:16}.  The electro-mechanical performance of the device can be characterized by measuring the RF (mag.) reflection ($|S_{11}|$) spectrum of the device with a vector network analyzer (VNA). A typical broadband scan is shown Fig.\ref{fig:2_S11}, with the insets showing narrowband scans of two modes.
The periodic series of dips in the $|S_{11}|$ spectrum correspond to the successive longitudinal acoustic eigenmodes of the structure. The 2D $z$-displacement profile (modelled using a finite element solver COMSOL) of three successive modes is shown in Fig.\ref{fig:1}(d), where the displacement profiles clearly show acoustic energy trapping due to mass loading from the Al electrode \cite{Shockley2005TrappedEnergyCrystals}. As the acoustic wave dissipation with frequency $\sim f^2$ \cite{auld1973acoustic}, having a short acoustic cavity ($L^{cav}_{mech} \approx 11.5\, {\mu}m$) allows us to increase the acoustic energy density (and the corresponding cavity displacement) at high frequencies. In fabricated devices, $Q_{mech}$ varies depending on the mode number and lies between 500-800, limited mainly by the surface roughness of the top Al electrode. The device geometry also serves as a planar optical Fabry-Perot (FP) cavity ($L^{cav}_{opt} \approx 10.5 {\mu}m$) with the aluminum contact serving as the top mirror. By depositing a thin film of gold (25-30 nm) on the bottom silicon side, we can improve the bottom mirror reflectivity and boost the optical quality factor ($Q_{sim}\,{\approx}\,800$, $Q_{expt}\,{\approx}\,400$). 
\begin{figure}[hbtp]
    \centering
    \includegraphics[width = \linewidth]{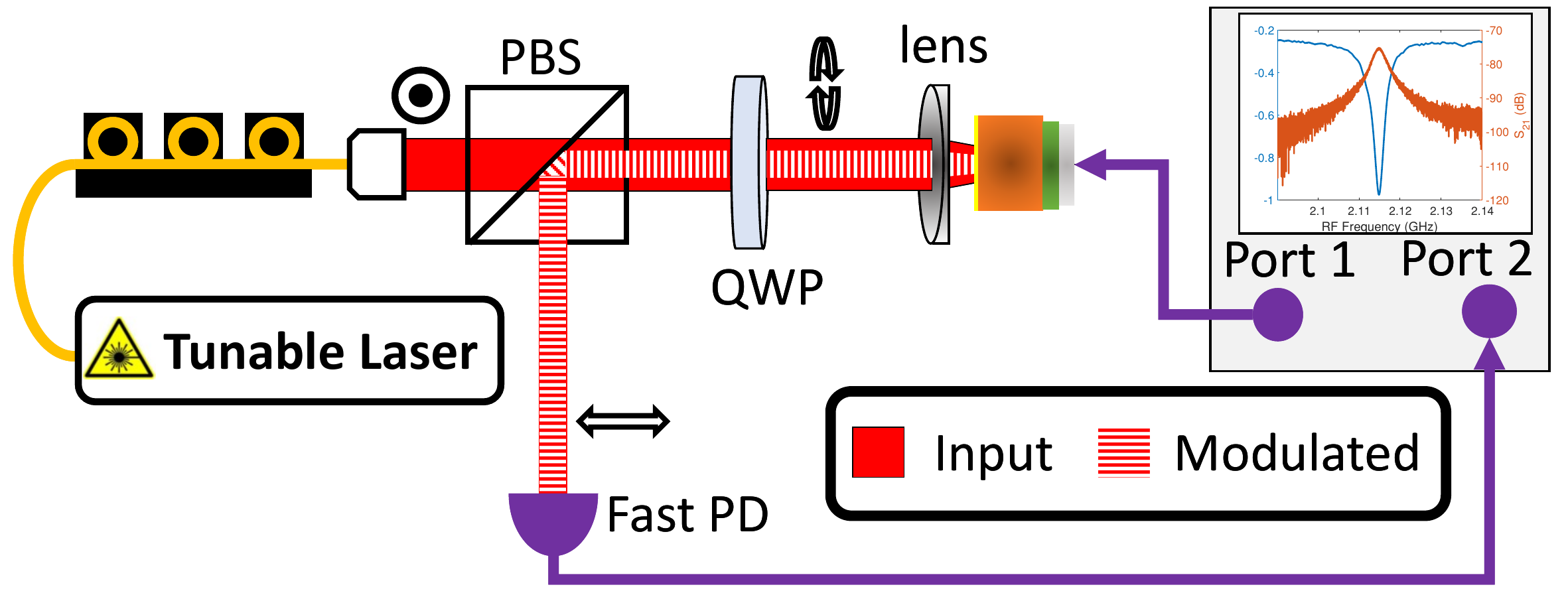}
    \caption{Experimental setup used for AO modulation measurement. The polarizing beam splitter (PBS) and quarter waveplater (QWP) are used to separate the modulated reflected beam from the incident beam.}
    \label{fig:3_setup}
\end{figure}\\
The AO performance of the device is characterized using the setup shown in Fig.\ref{fig:3_setup}, where the modulator is operated in reflection from the silicon (gold coated) side (Fig.\ref{fig:1}(c)). We use a polarizing beam splitter in conjunction with a quarter wave plate to separate the reflected beam from the incident beam. One of the optical FP modes measured using such a reflection scan is shown in Fig.\ref{fig:5} (green curve). For the AO modulation measurement, the laser wavelength is set at the maximum slope of the optical cavity reflection spectrum. This transforms the optical phase (frequency) fluctuations induced by the cavity displacement into amplitude fluctuations which can be detected with a high-speed photodetector (PD). An RF signal from the VNA is used to drive the cavity mechanics and the output of the PD, which measures the modulation, is fed back to the VNA. The transmission response (mag.) ($|S_{21}|$) of the VNA, thus measures the AO transfer function and can be used to extract the AO modulation index \cite{Balram2016CoherentCircuits}. A representative broadband scan of the $|S_{21}|$ spectrum is shown in Fig.\ref{fig:4_S11_S21}(a), overlaid with the $|S_{11}|$ spectrum. Narrowband scans around a couple of modes are shown in Fig.\ref{fig:4_S11_S21}(b-c) showing good correspondence between the $|S_{11}|$ and $|S_{21}|$ data.
\section{Discussion}
\begin{figure}[htbp]
    \centering
    \includegraphics[width = \linewidth]{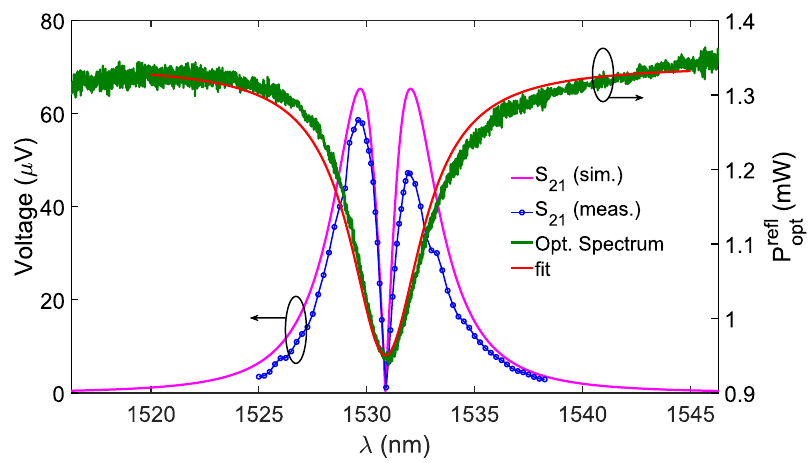}
    \caption{Experimental data (dark blue) and theoretical fit (magenta) for the AO modulation as a function of laser wavelength. The reflection spectrum of the optical cavity mode (green), along with Lorentzian fit (red) are also shown.}
    \label{fig:5}
\end{figure}

\begin{figure}[htb]
    \centering
    \includegraphics[width = \linewidth]{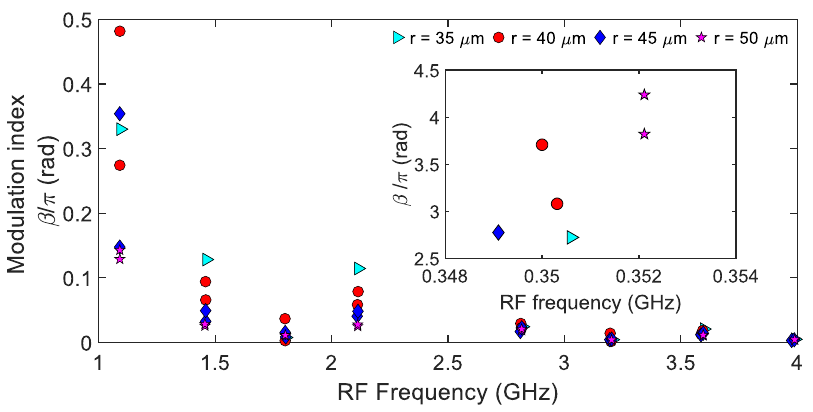}
    \caption{Modulation index ($\beta$) as function of mode frequency ($P_{RF}$ = 1 mW) for devices with varying disk radii ($r$). The $\beta$ for the low frequency mode is shown in the inset. The scatter among the data primarily arises from variable RF and optical performance in nominally identical devices due to fabrication imperfections.}
    \label{fig:6}
\end{figure}

\begin{figure}[hbtp]
    \centering
    \includegraphics[width = \linewidth]{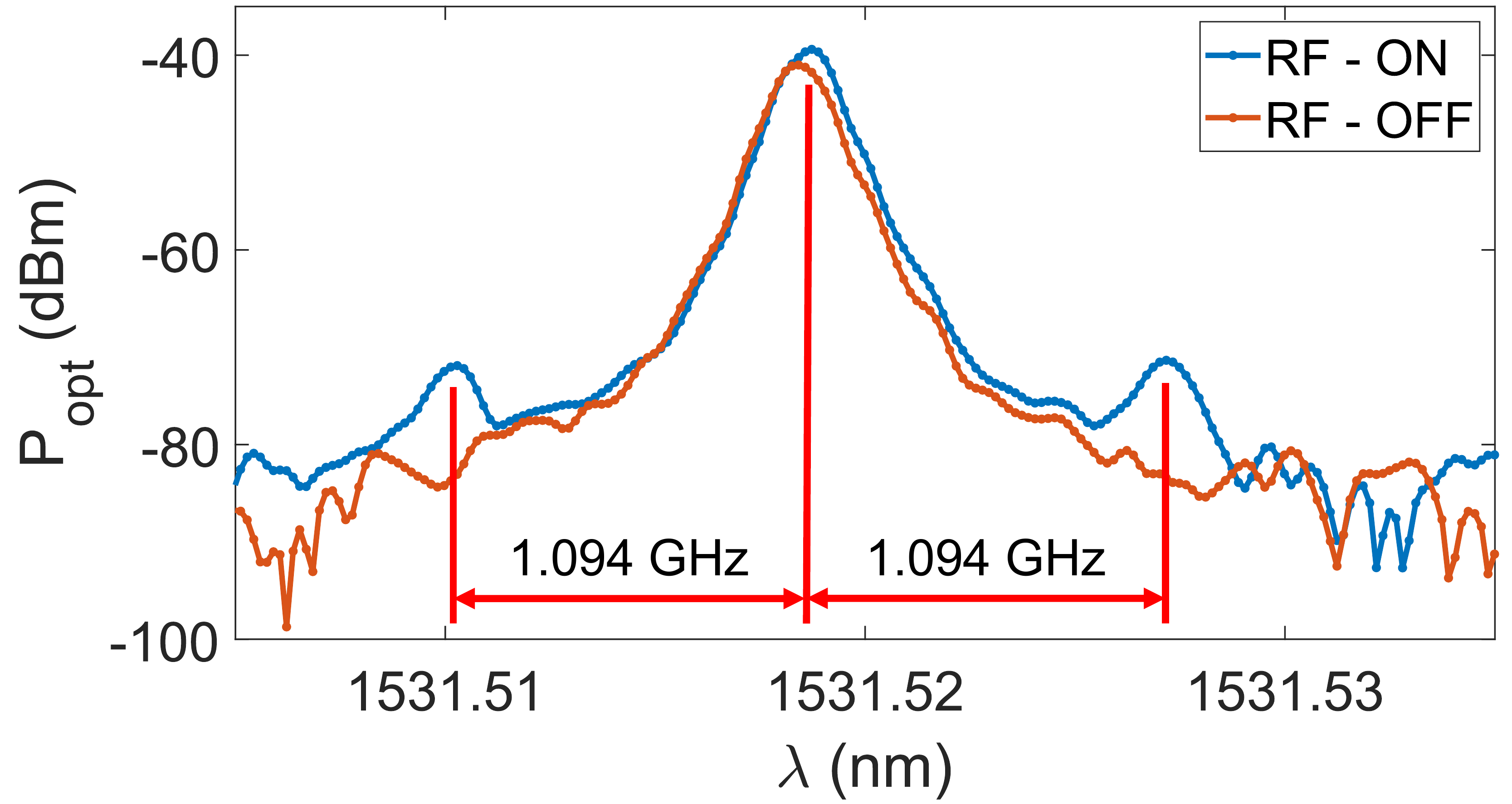}
    \caption{Optical reflection spectrum ($P_{RF}$ = 40 mW), with modulation ON (blue) and OFF (orange) with the pump at the optical resonant frequency of the cavity.}
    \label{fig:7_OSA}
\end{figure}

\begin{figure}[hbtp]
    \centering
    \includegraphics[width = \linewidth]{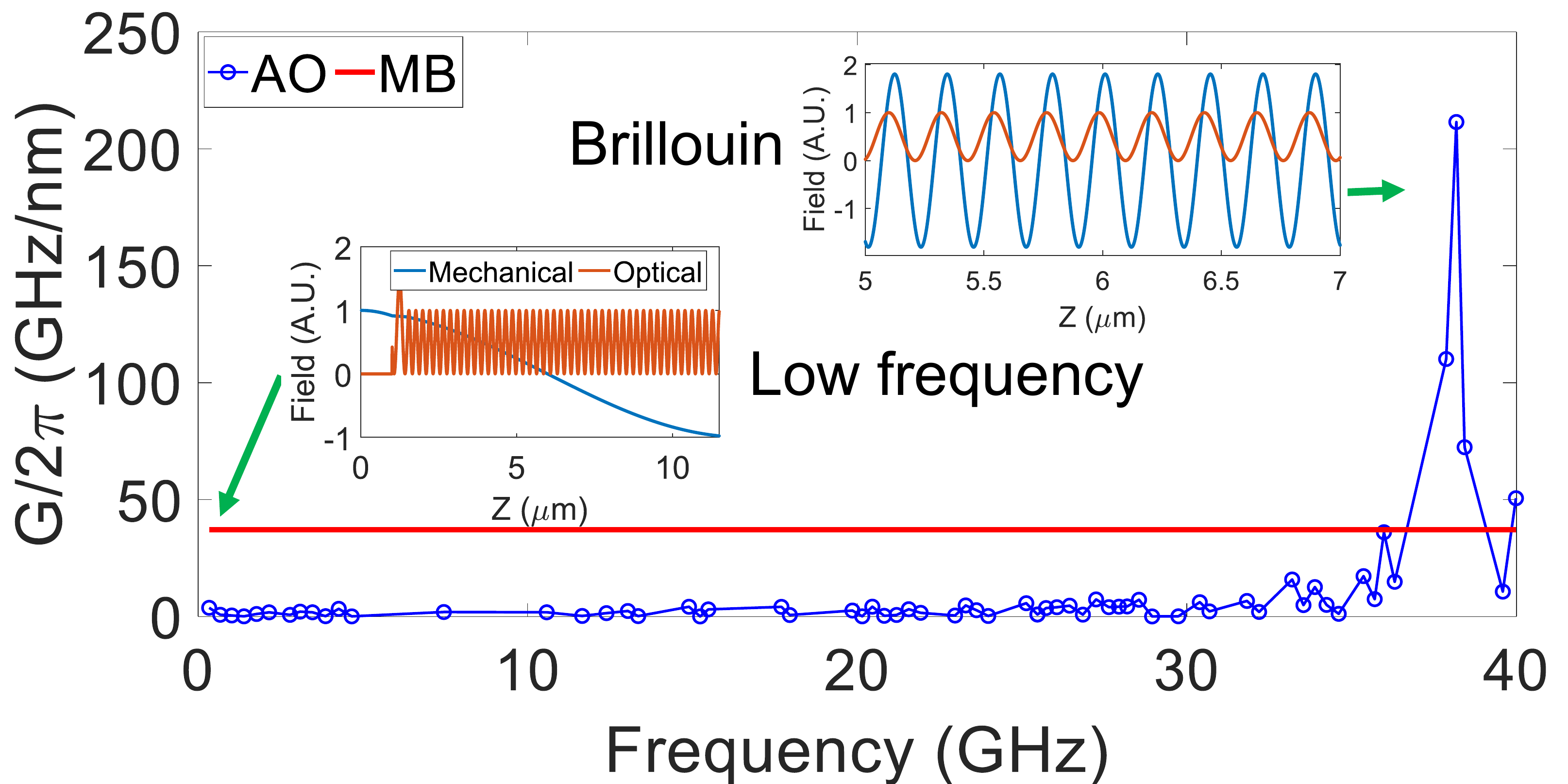}
    \caption{Optomechanical coupling ($G$) due to the moving-boundary (MB, red) and acousto-optic effects (AO, blue) for successive acoustic modes. The cavity acoustic and optical fields are shown at the two frequency extremes (low frequency and Brillouin) in the inset.}
    \label{fig:8_overlap}
\end{figure}

To quantify the AO interaction, we start with the coupled equations of cavity optomechanics, which characterize the interaction between the cavity optical and acoustic fields \cite{Gorodetsky2010DeterminationCalibration}:
\begin{equation}
    \dot{a}(t)=(i\Delta-iGz(t)-\kappa/2)a(t)+\sqrt{\eta_{c}\kappa}a_{in}(t)
\end{equation}
\begin{equation}
    \ddot{z}(t)+\Gamma\dot{z}(t)+\Omega_{m}^2z(t) = {\chi}V_{RF}(t)
\end{equation}
where the symbols represent: $a(t)$ cavity optical field, $\Delta$ laser detuning from the cavity resonance, $G=d\omega_{opt}/dz$ the optomechanical coupling rate which captures the shift in cavity frequency ($\omega_{opt}$) due to motion, $\eta_{c}$ cavity coupling coefficient, $\kappa$ the optical cavity decay rate, $a_{in}$ the input optical field, $z(t)$ the acoustic cavity displacement, $\Gamma$ the mechanical decay rate, $\Omega_{m}$ the resonant frequency of the cavity and ${\chi}V_{RF}(t)$ represents the piezoelectric driving term. The optical cavity frequency  oscillates at $\omega_{opt}(t) = \omega_{opt,0} + \beta \cos(\Omega_m t)$ where $\omega_{opt,0}$ is the unperturbed cavity frequency and  $\beta = G{\Delta}z/\Omega_{m}= \Delta f_m / f_m$ is the modulation index (ratio of peak frequency deviation $\Delta f_{m}$ to the modulation frequency $f_{m}$). The modulated cavity field can be expressed as:
\begin{equation}
\begin{split}
    a_{cav} = & a_{in}\sqrt{\eta_c k} \mathcal{L}(0)(R - \frac{i \beta \Omega_m \mathcal{L}(\Omega_m)}{2}e^{-i\Omega_m t} \\ & - \frac{i \beta \Omega_m \mathcal{L}(-\Omega_m)}{2}e^{+i\Omega_m t} )
    \label{eq:a_cav}
\end{split}
\end{equation}
where $R$ is the cavity reflectivity and $\mathcal{L}(\Omega)$ is the Lorentzian cavity response defined as:
\begin{equation}
\mathcal{L}(\Omega) = \frac{1}{\mathrm{-i}(\Delta + \Omega) + \kappa/2}
\label{eq:Lorentz}
\end{equation}
The field incident on the PD is $a_{out} = a_{in} - \sqrt{\eta_c \kappa}a_{cav}$, which provides a voltage output to the VNA:
\begin{equation}
    V_{out} = \eta_{PD}Z_{load}\left| a_{out} \right|^2
\end{equation}
where $\eta_{PD}$ is the PD responsivity (A/W) and $Z_{load}$ the VNA input impedance. This can be expressed in terms of the cavity parameters as
\begin{equation}
\begin{split}
    V_{out} = & 2 \Re \{ \mathrm{i} \eta_c \kappa P_{in} \Omega_m \frac{\beta}{2} ( R[\mathcal{L}_{S}^*\mathcal{L}_{0}^*-\mathcal{L}_{AS}\mathcal{L}_{0}] \\ & +\eta_c \kappa |\mathcal{L}_0|^2 \frac{\beta}{2}\left[\mathcal{L}_{AS}-\mathcal{L}_{S}^*\right] ) \} \eta_{PD} Z_{load}
    \label{eq:V_out_long}
\end{split}
\end{equation}
where $P_{in}$ is the input optical power and $\mathcal{L}_i$ with $i=\{S,AS,0\}$ is the Lorentzian function in Eq.\ref{eq:Lorentz} with detuning corresponding to the Stokes ($-\Omega_{m}$), Anti-Stokes ($+\Omega_{m}$), and pump fields ($0$). To verify the validity of our model, we scanned the laser across the optical resonance and measured the AO modulation ($|S_{21}|$ amplitude) as a function of laser wavelength for a fixed mechanical resonance ($f_m = 1.094$ GHz). Fig.\ref{fig:5} plots the measured data (dark blue), along with with the theoretical fit from eq.\ref{eq:V_out_long} (magenta) showing good agreement.
\\
The modulation index ($\beta$) for each mode can be extracted from the broadband $|S_{21}|$ scan (Fig.\ref{fig:4_S11_S21}(a)) and the results are shown in Fig.\ref{fig:6} for disks of different radii ($r: 35-50 {\mu}m$).
\\
We measured a maximum $\beta$ $\approx\pi/2$ for the $\approx$ 1.096 GHz resonance and $\approx \pi/8$ for the $\approx$ 2.115 GHz resonance, for an RF input power ($P_{RF}$) of 1 mW. The highest $\beta$ ($\approx 4.5 \pi$) is achieved for the low frequency $\approx350$ MHz mode (shown in the inset of Fig.\ref{fig:6}). This is expected as the low frequency mode has a higher displacement amplitude (${\Delta}L^{cav}_{mech}\approx 125$ pm) and correspondingly larger cavity frequency shift ($\propto G{\Delta}L^{cav}_{mech}$).Although we achieve large phase modulation, the amplitude modulation in the reflected optical signal is small on account of the low optical quality factor ($Q_{opt}\approx400$). From eq.\ref{eq:a_cav}, we can see that the sideband amplitude scales $\sim \beta\Omega_m/\kappa$ which limits the scattering efficiency in the regime $\Omega_{m} \ll \kappa$, which is the case here. In order to measure the power in the sidebands directly (complementary to the homodyne VNA $|S_{21}|$ measurement), we added an RF amplifier (16 dB gain) to enhance $\beta$ and observe the modulated optical spectrum, shown in Fig.\ref{fig:7_OSA}.
\\
The AO coupling is dominated by the moving boundary (MB) contribution, which can be approximated as $G\approx 2\omega_{opt}/L^{cav}_{opt}$ for the even (symmetric) displacement modes (e.g. the 1.09 and 1.78 GHz modes in Fig.\ref{fig:1}(d)) and 0 for the odd (anti-symmetric) displacement (1.45 GHz mode in  Fig.\ref{fig:1}(d)). The AO contribution can be estimated (in a 1D dissipation-free approximation) from $\int S_{zz}(z)p_{12}|E(z)|^2dz$ where $S_{zz}(z)$ is the strain, $p_{12}$ is the photoelastic coefficient and $E(z)$ the electric field of the optical cavity mode. The relative contributions of the two effects, for each of the acoustic cavity resonances, are plotted in Fig.\ref{fig:8_overlap}. The MB contribution dominates at all frequencies, except near the Brillouin frequency of silicon, where the AO interaction becomes phase-matched ($\vec{k}_{acoustic}=2\vec{k}_{opt}$) \cite{Renninger2018BulkOptomechanics,voloshinov2018design} and the optomechanical coupling can be significantly enhanced. 
\\
The even-odd mode coupling can be clearly discerned in the high frequency modes (> 2.5 GHz) in Fig.\ref{fig:4_S11_S21}(a) (green and red arrows) where the modes alternate between finite and zero modulation. We are not able to observe this trend in the low frequency modes (< 2 GHz) and believe this is primarily due to mode coupling, owing to the highly symmetric (circular) geometry of our electrodes. By shaping our electrodes appropriately, we will address this issue in our next fabrication run. Moving forward, we hope to improve the viability of this approach by focusing our research efforts on three fronts: improving $Q_{opt}$ by enhancing the bottom mirror reflectivity, increasing the modulation frequency ($\approx \, 10 $ GHz) by improving transducer design and scaling the architecture to 2D arrays.

\section*{Funding Information}
Europractice (MEMS First-User Stimulus), ERC (SBS3-5 758843) and EPSRC (EP/N015126/1)
\section*{Acknowledgments}
K.C.B. would like to thank D. Bode, H. Gersen, L. Kling, G. Marshall, J. Rarity, K. Srinivasan and M. Wu for valuable discussions and suggestions. 

\bibliography{OL_arxiv}% Produces the bibliography via BibTeX.

\end{document}